\documentclass[twocolumn,amsmath,showkeys,amssymb,superscriptaddress,longbibliography,floatfix, pre]{revtex4-2}

\pdfoutput=1

\usepackage{bm}
\usepackage{yfonts}
\usepackage{graphicx,epsfig}
\usepackage{amsmath,amssymb,mathtools}
\usepackage[rgb]{xcolor}
\usepackage{hyperref}
\usepackage{braket}
\def\be{\begin{equation}}
\def\ee{\end{equation}}

\def\bc{\begin{center}}
\def\ec{\end{center}}
\def\bea{\begin{eqnarray}}
\def\eea{\end{eqnarray}}

\newcommand{\Avg}[1]{\left\langle{#1}\right\rangle}

\begin{document}

\title{Beyond holography:\\ the entropic quantum gravity  foundations of anisotropic diffusion}

\author{Ginestra Bianconi}
\email{ginestra.bianconi@gmail.com}
\affiliation{School of Mathematical Sciences, Queen Mary University of London, London, E1 4NS, United Kingdom}

\begin{abstract}
Recently, thanks to the development of artificial intelligence (AI) there is increasing scientific attention in establishing the connections between theoretical physics and AI. Traditionally, these connections have been focusing mostly on the relation between string theory and image processing and involve important theoretical paradigms such as holography.    Recently G. Bianconi has formulated the Gravity from Entropy (GfE) approach to quantum gravity in which gravity is derived from the geometric quantum relative entropy (GQRE) between two metrics associated with the Lorentzian spacetime. Here  it is demonstrated that the famous Perona-Malik algorithm for image processing is the gradient flow that maximizes  the GfE action in its simple warm-up scenario. Specifically, this algorithm is the outcome of the maximization of the GfE action calculated   between two Euclidean metrics: the one of the support of the image and the one induced by the image. As the Perona-Malik algorithm is known to preserve sharp contours, this implies that  the GfE action, does not in general lead to uniform images upon iteration of the gradient flow dynamics as it would be intuitively expected from entropic actions maximising  classical entropies. Rather, the outcome of the maximization of the GfE action is compatible with the preservation of complex structures. These results provide the geometrical and information theory foundations for the Perona-Malik algorithm   and might contribute to  establish deeper connections between GfE, machine learning and brain research.
\end{abstract}
\maketitle

Recently there is an increasing recognition of the common mathematical foundations of theoretical physics and  artificial intelligence (AI) algorithms \cite{levine2024machine,carleo2019machine,carifio2017machine,he2023deep,carrasquilla2017machine,mendes2024wave,millan2025topology,wang2024dirac}.
Specifically there is a growing consensus on the fundamental role of topology and geometry to inform the most recent developments of AI and network theory. This scientific interest is currently leading to the fast developments of very vibrant research fields such as   topological and geometrical  machine learning~\cite{papamarkou2024position,bronstein2017geometric} and to topological higher-order network dynamics~\cite{bianconi2021higher,battiston2021physics,millan2025topology} that might lead to  a more comprehensive understanding of brain dynamics~\cite{petri2014homological,reimann2017cliques,santoro2023higher,faskowitz2022edges,barabasi2023neuroscience}.
At the interface between problems in brain research an in AI, a key research question is whether theoretical physics can inspire geometric diffusion models~\cite{di2022understanding,bronstein2021geometric}
 that can  be crucial to develop the foundation of  the next generation of unsupervised or self-supervised learning algorithms~\cite{schmarje2021survey}. 
In addition to geometry and topology,  information theory is also recognized as  fundamental for both AI \cite{moore2011nature,mezard2009information} and brain research \cite{friston2010free}. In particular,  the relative entropy, also known as Kullback-Leibler entropy,  is acquiring a fundamental role in AI and has given rise to very successful theoretical concepts and algorithms as the information bottleneck principle~\cite{tishby2000information} and diffusion models \cite{yang2023diffusion,chamberlain2021grand}.
 {In this work we define the geometric quantum relative entropy  (GQRE) in Euclidean spaces and we show that the GQRE  provides  solid information theory foundations to  geometric anisotropic diffusion algorithms. By doing so, we  establish the connection between these anisotropic diffusion algorithms and  the recently proposed {\em Gravity from entropy} (GfE) approach to quantum gravity \cite{bianconi2025gravity}.}

The relation between artificial intelligence,  network theory  and theoretical physics has a long history.  Theoretical physics  inspiration from computer vision has led to important conceptual frameworks such as  the formulation of the holographic principle~\cite{hooft2001holographic,susskind1995world,maldacena2005illusion}. 
The holographic principle was originally motivated~\cite{hooft2001holographic} to obtain the area law for the entropy of black-holes. 
More in general, this principle states that our three-dimensional universe might be encoded in a two dimensional surface as an hologram and leading to a very vibrant and active research direction in theoretical physics~\cite{susskind1995world,maldacena1999large}.

On the other side, important proposals to reconcile image processing algorithms and in general AI algorithm with theoretical physics are at the forefront of physics-inspired AI \cite{di2022understanding,bronstein2021geometric}.
The root of this field can be traced back to  the work \cite{kimmel1997high} by  Sochen, Kimmel and Malladi.  {This work is the first work that has framed image processing into a differential geometry theory involving the metric induced by the image. Moreover this work has proposed the adoption of the  Polyakov action of string theory in order to  formulate  modified anisotropic diffusion models for image processing. }

The cross-fertilization of ideas between theoretical physics and specifically quantum information and network science is also very fertile \cite{passerini2008neumann,anand2009entropy,anand2011shannon,de2016spectral,villegas2023laplacian,de2015structural,morone2020fibration}.
In this context,  quantum entropy has found applications  in the characterization of complex network structure, thanks to the use of the Von Neumann entropy associated with the graph Laplacian. This so-called Von Neumann entropy of networks  has been originally proposed in Ref.~\cite{passerini2008neumann} by  Passerini and Severini and since then has found wide applications in the theory of simple and multilayer networks~\cite{anand2009entropy,anand2011shannon,
de2016spectral,de2015structural} and provides the underlying theoretical framework for the recently proposed renormalization group approach in network theory~\cite{villegas2023laplacian}. However, the quantum relative entropy that is so central in quantum information theory ~\cite{vedral2002role} has not  yet been recognized to have wide applications in either AI or network science.

 {Recently, in Ref. \cite{bianconi2025gravity}, the author has combined quantum mechanics with gravity in a comprehensive statistical mechanics and quantum information theory framework known as {\em Gravity from Entropy} (GfE). Specifically she has formulated a geometrical definition of quantum relative entropy called GQRE in Lorentzian spacetimes.  In this framework the metrics associated with spacetime are treated as quantum operators and the GfE action  is determined by a Lagrangian given by the GQRE between the metric of the manifold and the metric induced by the matter field and curvature. Thus, the action of GfE is a quantum information theory action that fully captures the interplay between geometry and the  matter fields defined on the manifold.} Interestingly,  the GQRE can  also be calculated for the Schwarzschild black hole giving rise to an area law for large Schwarzschild radius~\cite{bianconi2025quantum} without obeying the holographic principle.

 {Here, we demonstrate that the action of GfE,  is key for geometric anisotropic diffusion models. In particular we reveal that in one of its simplest incarnations, the GfE action,  associated to a GQRE Lagrangian  defined on Euclidean spaces,  provides the quantum information theory foundations for the famous  Perona-Malik algorithm~~\cite{perona1990scale} for image processing.  }
The  Perona-Malik algorithm  is  a fundamental reference model in the field of image processing~\cite{kichenassamy1997perona,guidotti2009some,buades2005review} that denoises an image by implementing a diffusion process. This diffusion is however anisotropic and takes place in presence of a metric that is chosen to have an ad hoc functional form dependent on the contrast of the image. 
Previous attempts to relate the Perona-Malik algorithm to string theories and in particular the Polyakov action, \cite{kimmel1997high} have not been  able to justify the empirical choice of the functional form of the metric assumed in the Perona-Malik algorithm.
 {Here we provide solid quantum information theory foundations for  the Perona-Malik algorithm by defining the GQRE in an Euclidean setting and showing that  the Perona-Malik algorithm is given by the gradient flow that maximizes the GfE action proposed in Ref. \cite{bianconi2025gravity}.}
The  definition of the GQRE is here shown  to be rooted in the theory of von Neumann algebras~
\cite{witten2018aps,araki1999mathematical} and to constitute a significant advance with respect to the Araki entropy \cite{araki1975relative,ohya2004quantum}.
  {The GQRE provides an information theory quantity that is built from two metrics associated to the image: the true metric of the 2D flat Euclidean space that offers the support to the image, and the metric induced by the image defined on it.
Anisotropic diffusion emerges from the tension between these two metrics to reduce their differences quantified through the GQRE Lagrangian of GfE.
As the Perona-Malik is known to preserve sharp contours of the image, the GfE foundation of the Perona-Malik action also points out an important property of the GfE action. Indeed,  the maximization of the GfE action is compatible with the preservation of structure and complexity of the image. This is a notable feature of the GfE action as it reveals that, contrary to  the  expectations in scenarios when the classical entropy is maximized, the maximization  of the GfE action, based on the GQRE Lagrangian,  is compatible with non-homogeneous and complex outcomes.}

\begin{figure*}[htb]
  \includegraphics[width=1.55\columnwidth]{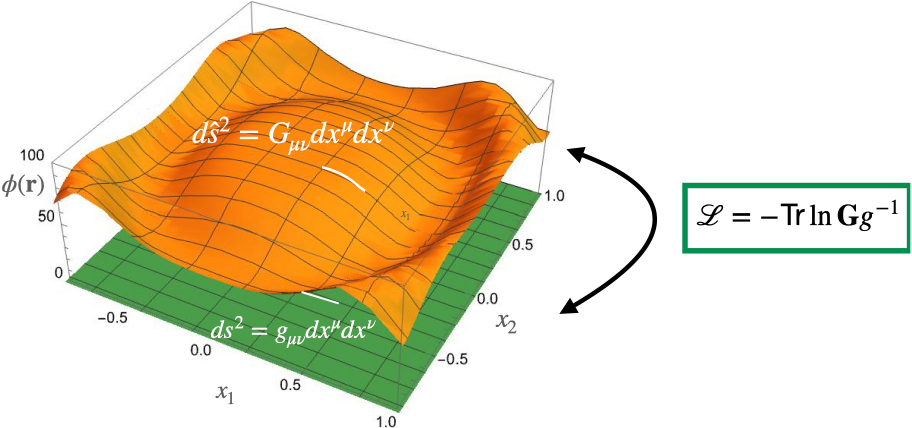}
  \caption{ {Schematic representation of the metrics associated to a black-and-white image and the GQRE Lagrangian of the GfE action which provides the foundations of the Perona-Malik algorithm. The flat $2D$ Euclidean manifold $\Omega$ (in green) that offers the support of the image, has points ${\bf r}$ of coordinates ${\bf r}=(x_1,x_2)$ and an Euclidean metric $g_{\mu\nu}=\eta_{\mu\nu}$. The infinitesimal distance $ds$ between points in $\Omega$ is defined by this metric and obeys $ds^2=g_{\mu\nu}dx^{\mu}dx^{\nu}$.  The image is associated to the surface $\mathcal{K}$  (in orange) embedded in a  flat 3D Euclidean metric,thus every point of $\mathcal{K}$ has coordinates $({\bf r}, \phi({\bf r}))$. 
The infinitesimal distance $d\hat{s}$ between two points  in this surface obeys  $d\hat{s}^2=G_{\mu\nu}dx^{\mu}dx^{\nu}$ where $G_{\mu\nu}$, given by Eq.(\ref{induced}), is the symmetric rank $2$ tensor that defines the metric induced by the surface $\mathcal{K}$ on the $2D$ flat support of the image $\Omega$. 
  The GfE action, associated to the Lagrangian $\mathcal{L}$, is given by the  GQRE between the metrics ${\bf G}$ and $g$. In this work we show  that the Perona-Malik algorithm is the gradient flow that maximizes  the GfE action.  Therefore the  Perona-Malik algorithm  emerges from the principle of maximization of the GfE action calculated between the metric induced by the image ${\bf G}_{\mu\nu}$ and the metric $g_{\mu\nu}$ of the 2D image support.}}
  \label{fig1}
 \end{figure*}

{\bf The Perona-Malik algorithm.}
We consider a $2D$ flat Euclidean manifold $\Omega$ of coordinates ${\bf r}=(x_1,x_2)\in \Omega$ with metric $g_{\mu\nu}=\eta_{\mu\nu}$ with $\eta_{\mu\nu}=1$ if $\mu=\nu$ and $\eta_{\mu\nu}=0$ otherwise.
The infinitesimal distance $ds$ between points in $2D$ is defined by this metric and obeys $ds^2=g_{\mu\nu}dx^{\mu}dx^{\nu}$.  
Note that the metric $g_{\mu\nu}$ and its inverse $g^{\mu\nu}$ will be central to transform vectors in one-forms and vice versa by lowering or raising the indices, i.e.
\bea
g_{\mu\nu}V^{\nu}=V_{\mu},\quad g^{\mu\nu}V_{\nu}=V^{\mu}.
\eea

On top of this manifold we define a function $\phi({\bf r})\in \mathbb{R}$ indicating the intensity of the colour of the single colour, black-white image in our simple setting.
Given an initial noisy image determined by the function $\psi({\bf r})$ the Perona-Malik algorithm~\cite{perona1990scale,guidotti2009some,buades2005review} proposes to reconstruct the true image by performing a Laplace-Beltrami diffusion with metric. Specifically, the reconstructed image $\phi({\bf r})$ is found by integrating the system
\bea
\frac{d\phi({\bf r},t)}{dt}&=&\nabla_{\mu}\rho(|\nabla\phi|^2)\nabla^{\mu}\phi({\bf r},t)\nonumber\\
\phi({\bf r},0)&=&\psi({\bf r}).
\label{PM}
\eea
where the metric $\rho(|\nabla\phi|^2)$ is taken to be
\bea
\rho(|\nabla\phi|^2)=\frac{1}{1+\alpha |\nabla\phi|^2},
\label{rho}
\eea
and where $\alpha\in \mathbb{R}^+$ is a parameter of the model. The power and beauty of this model is that the metric evolves together with the diffusion model and the reconstruction of the image. It is clearly highly desirable to derive  the specific functional form of the metric  $\rho(|\nabla\phi|^2)$ in a principled way. However, in the original work of Perona and Malik \cite{perona1990scale} there are no fundamental information theory principle driving this choice which remains an {\em ad hoc} choice in order to achieve anisotropic diffusion and a good performance of the algorithm.   {Additionally, also the string theory approach to anisotropic diffusions proposed in Ref. \cite{kimmel1997high} does not provide this explanation.}

As we will see in this article, this particular choice of the metric  $\rho(|\nabla\phi|^2)$ is exactly what is predicted by the GfE action.

{\bf Induced metric.}
The  set of points 
$({\bf r},\phi({\bf r}))$ with ${\bf r}\in \Omega$  
defines a $2D$ surface $\mathcal{K}$ immersed in $3D$ (see Figure 1). 
At any given point of $\mathcal{K}$, the tangent vectors are given by ${\bf e}_1=(1,0,\nabla_{x^1}\phi)$ and ${\bf e}_2=(0,1,\nabla_{x^2}\phi)$. Let us assume that the 3D embedding space of $\mathcal{K}$ has flat Euclidean metric with diagonal elements $(1,1,\alpha)$, where $\alpha$ is a positive real constant.  The infinitesimal distance $d\hat{s}$ between points ${\bf X}$ and ${\bf X}+\delta{\bf X}$ with $\delta {\bf X}={\bf e}_1 dx^1+{\bf e}_2 dx^2$ in $\mathcal{K}$ obeys  $d\hat{s}^2=G_{\mu\nu}dx^{\mu}dx^{\nu}$ where ${G}_{\mu\nu}$ are the elements of the real and symmetric rank $2$ tensor that defines  the induced metric on $\Omega$.
Specifically, we have that the induced metric ${\bf G}$ on the 2D manifold $\Omega$ that provides the support of the image, is given by 
\bea
{ G}_{\mu\nu}=g_{\mu\nu}+\alpha \nabla_{\mu}\phi\nabla_{\nu}\phi.
\label{induced}
\eea

Thus the considered manifold $\Omega$ is associated with two metrics; the metric $g_{\mu\nu}$ and the metric $G_{\mu\nu}$.
Note that in the following we will use $\hat{\bf G}_{\mu\nu}$ when referring to either one of these two metrics.
 
According to the GfE approach  here we will  treat both $g$ and ${\bf G}$ as quantum operators and we will consider the action $\mathcal{S}$ given by the GQRE between these two metrics. We will show that the Perona-Malik algorithm can be obtained as the gradient flow that maximizes the GfE action.\\

{\bf Eigenvalues of the metrics associated to the manifold.}
The GQRE will be defined in terms of the eigenvalues of the true metric $g$ and the induced metric ${\bf G}$.
In order to define  the eigenvalues and eigenvectors of these two metrics in a rotationally invariant way, we define $\lambda$ as an eigenvalue of $\hat{\bf G}_{\mu\nu}$ if it solves the eigenvalue problem
\bea
\hat{G}_{\mu\nu}[V^{(\lambda)}]^{\nu}=\lambda V^{(\lambda)}_{\mu}.
\label{eig1}
\eea
Thus this eigenvalue problem is the usual eigenvalue problem for the matrix $\hat{\bf G}g^{-1}$ as the above equation reduces to
\bea
\hat{\bf G}_{\mu\nu}g^{\nu\rho}V_{\rho}=\lambda V_{\mu}.
\eea
It follows that all the eigenvalues $\lambda^{\prime}$ of $g_{\mu\nu}$ are  equal to one $\lambda^{\prime}_n=1$, for $n\in \{1,2\}$ independently of the choice of the metric $g_{\mu\nu}$
Instead, the eigenvalues $\lambda$ and the associated (non-normalized) eigenvectors $V_{\mu}$ of the induced metrics ${\bf G}_{\mu\nu}$,  are given by
\bea
\lambda_1=(1+\alpha|\nabla \phi|^2),\quad V_{\mu}=\nabla_{\mu}\phi\nonumber \\
\lambda_2=1,\quad \left(\eta_{\mu\nu}-\frac{\nabla_{\mu}\phi\nabla_{\nu}\phi}{|\nabla\phi|^2}\right)V^{\nu}.
\label{lambda2}
\eea
An action that is only dependent on these eigenvalues will be clearly rotational invariant.\\
{\bf The GfE action.}  {The GfE action \cite{bianconi2025gravity}  is associated with a Lagrangian given by the GQRE between two Lorentzian metrics: the metric of the manifold  and the metric induced by the matter fields, and curvature. In the context of the Perona-Malik algorithm we consider the warm-up scenario of GfE where the two metrics associated to the GQRE are Euclidean of the metric induced by the matter-field and curvature is played by ${\bf G}_{\mu\nu}$ which is defined in Eq.(\ref{induced})  and schematically described in Figure 1. }
 {The action $\mathcal{S}$ of GfE is given by 
\bea
\mathcal{S}=\frac{1}{2}\int_{\Omega} \sqrt{|-g|}\mathcal{L} d{\bf r},
\label{action}
\eea
where the Lagrangian $\mathcal{L}$ is given by the GQRE between $g$ and ${\bf G}$. 
Anticipating the results of the next paragraph,  the GQRE Lagrangian $\mathcal{L}$ in Euclidean space that encodes for the Perona-Malik algorithm is given by 
\bea
\mathcal{L}=-\mbox{Tr}\ln {\bf G}g^{-1}.
\label{Lag00}
\eea
This definition of the GQRE implies that  the GQRE that encodes for the Perona-Malik algorithm, can be also  expressed in terms of the eigenvalues $\lambda_n$ of ${\bf G}$ and the eigenvalues $\lambda^{\prime}_n$ of $g$ in a  familiar form for the relative entropy. Indeed  we have
\bea
\mathcal{L}=-\sum_{n=1}^2\ln \lambda_n=\sum_{n=1}^2\lambda^{\prime}_n(\ln \lambda^{\prime}_n-\ln \lambda_n),
\eea
where we have used $\lambda_n^{\prime}=1$ for  $n\in \{1,2\}$.
Using the explicit expression of $\lambda_n$ given by Eq.(\ref{lambda2}), we
obtain
\bea
\mathcal{L}=-\ln(1+\alpha|\nabla\phi|^2),
\label{Lag}
\eea
which constitute the Euclidean version of the warm-up scenario of GfE proposed in Ref.\cite{bianconi2025gravity}. }

{\bf First principles derivation of the  GQRE in Euclidean space Eq. (\ref{Lag00}) and its connection with Araki entropy.}  {In this paragraph we formulate the GQRE in Euclidean space. We discuss the quantum information theory foundations  of the  GQRE  Lagragian $\mathcal{L}$ (Eq.(\ref{Lag00}))   and we relate it to the Araki entropy \cite{araki1975relative,ohya2004quantum}.
Consistently with the GfE approach \cite{bianconi2025gravity}, here we consider a Dirac-K\"ahler  \cite{kruglov2002dirac,becher1982dirac}  interpretation of the gradient in which $\nabla_{\mu}\phi$ describes the component of a one-form. 
Therefore, in  order to provide the theoretical foundations for our definition of the GQRE, 
we will  consider  Hilbert spaces formed by the direct sum of a zero-form and a one-form. }
 The generic vector  $\ket{\Psi}$ of the considered Hilbert space $\mathcal{H}$ is given by 
\bea
\ket{\Psi}={{\phi}}\oplus {\omega}_{\mu} dx^{\mu},\label{phi1}
\eea
 {Thus both the image (encoded by $\phi\oplus 0_{\mu}dx^{\mu}$) and the gradient of the image encoded by $0\oplus\nabla_{\mu}\phi dx^{\mu}$  can be interpreted as elements of this Hilbert space.}
The scalar product between $\ket{\Phi}$ and  another generic vector $\ket{\Phi}$  given by 
$\ket{\Phi}=\hat{\phi}\oplus \hat{\omega}_{\mu} dx^{\mu}$
is defined as 
\bea
\Avg{\langle \Psi,\Phi\rangle}=\int \sqrt{-|g|} \Big(\bar{\phi}\hat{\phi}+\bar{\omega}_{\mu}\hat{\omega}^{\mu}\Big) d{\bf r},
\eea
where $\hat\omega^{\mu}=g^{\mu\rho}\hat\omega_{\rho}$ .
Thus the metric tensor $\tilde{g}^{-1}$ associated to this scalar product is given by 
\bea
\tilde{g}^{-1}&=&1\oplus g^{\mu\nu}dx_{\mu}\otimes dx_{\nu}.
\eea

All vectors  $\ket{\Phi}$ in the Hilbert space $\mathcal{H}$ must satisfy
\bea
\Avg{\bra{\Phi}\Phi\rangle}<\infty.
\label{hilb}
\eea 

Starting from the induced metric ${\bf G}$ we can construct the topological induced metric  $\tilde{\bf G}$  given by 
\bea
\tilde{\bf G}&=&1\oplus {G}_{\mu\nu}dx^{\mu}\otimes dx^{\nu}.
\eea
This topological induced metric can be interpreted as a  quantum operator $\tilde{\bf G}:\mathcal{H}\to \mathcal{H}$ where  $\tilde{\bf G}\cdot \ket{\Phi}\in \mathcal{H}$,  
\bea
\tilde{\bf G}\cdot \ket{\Phi}&=&\phi\oplus {G}_{\mu\nu}\omega^{\nu}dx^{\mu}.
\eea

The metric $\tilde{g}$ of the manifold $\Omega$ can be used to define a dual Hilbert space $\mathcal{H}^{\star}$. To this end we define the dual of $\ket{\Psi}$ as $\ket{\Phi^{\star}}$ and the dual of $\ket{\Phi}$ as $\ket{\Phi^{\star}}$ given by
\bea
\ket{\Psi^{\star}}={{\phi}}\oplus {\omega}^{\mu} dx_{\mu},\quad
\ket{\Phi^{\star}}=\hat{\phi}\oplus \hat{\omega}^{\mu} dx_{\mu},
\eea
where $\omega^{\mu}=g^{\mu\rho}\omega_{\rho},\hat{\omega}^{\mu}=g^{\mu\rho}\hat{\omega}_{\rho}$.

The  scalar product $\Avg{\Avg{ \Psi^{\star},\Phi^{\star}}}_{\star}$ is mediated by $\tilde{g}$ 
given by 
\bea
\tilde{g}=1\oplus g_{\mu\nu}dx^{\mu}\otimes dx^{\nu}
\eea
and satisfies 
\bea
\Avg{\langle{\Psi},\Phi\rangle}=\Avg{\langle \Psi^{\star},\Phi^{\star}\rangle}_{\star}.
\label{Hdual}
\eea 

The dual operator $\tilde{\bf G}^{\star}:\mathcal{H}^{\star}\to\mathcal{H}^{\star}$ of $\tilde{\bf G}$  is given by 
\bea
\tilde{\bf G}^{\star}&=&1\oplus {[G^{\star}]}^{\mu\nu}dx_{\mu}\otimes dx_{\nu}.
\eea
which must satisfy
\bea
\Avg{\Avg{{\Psi},\tilde{\bf G}\cdot\Phi}}=\Avg{\Avg{ \tilde{\bf G}^{\star}\cdot \Psi^{\star}, \Phi^{\star}}}_{\star}.
\label{HdualG}
\eea 
for any arbitrary choices of $\ket{\Psi}$ and $\ket{\Phi}$.
Here the action of $\tilde{\bf G}^{\star}$ on the generic dual vector $\bra{\Phi^{\star}}$ obeys
\bea
\tilde{\bf G}^{\star}\cdot \ket{\Phi^{\star}}&=&\phi\oplus {[G_{(1)}^{\star}]}^{\mu\nu}\omega_{\nu}dx_{\mu}.
\eea
Thus, using the symmetry of $G_{\mu\nu}$ we  conclude that $\tilde{\bf G}^{\star}$ is related to $\tilde{\bf G}$ by
\bea
\tilde{\bf G}^{\star}=\tilde{g}^{-1}\tilde{\bf G}\tilde{g}^{-1}.
\eea
indicating that 
\bea
{[G^{\star}]}^{\mu\nu}&=&g^{\mu\rho}{G}_{\rho\sigma}g^{\nu\sigma}.
\eea
 
The topological metrics  $\tilde{\bf G}$ and $\tilde{\bf G}^{\star}$ that we consider in this work are respectively elements of  the algebras $\textswab{U}$ and $\textswab{U}^*$ that generalizes the $C^*$ algebra \cite{araki1999mathematical}. The norm associated to the topological metric $\tilde{\bf G}\in \textswab{U}$ is   equal to the norm associated to its dual, i.e.. $\|\tilde{\bf G}\|=\|\tilde{\bf G}^{\star}\|$ with
\bea
\|\tilde{\bf G}\|=\|\tilde{\bf G}^{\star}\|=\int \sqrt{|-g|}\mbox{Tr}_F \Big(\tilde{\bf G}\tilde{\bf G}^{\star}\Big) d{\bf r},
\label{norm}
\eea
where $\mbox{Tr}_F \Big(\tilde{\bf G}\tilde{\bf G}^{\star}\Big)$ is given by  
\bea
\mbox{Tr}_F \Big(\tilde{\bf G}\tilde{\bf G}^{\star}\Big)&=&1+{G}_{\mu\nu}{G}^{\nu\mu}.
\eea

For a topological metric $\tilde{\bf G}$  interpreted as a quantum operator in $\textswab{U}$ we define the square root of the modular operator $\bm\Delta^{1/2}_{\tilde{\bf G},g}:\mathcal{H}\to \mathcal{H}$ as 
\bea
\bm\Delta^{1/2}_{\tilde{\bf G},g}=\sqrt{\tilde{\bf G}\tilde{\bf G}^{\star}}=\tilde{\bf G}\tilde{g}^{-1},
\eea
where the last identity is derived under the assumption that $\tilde{\bf G}$ is positively definite, i.e. it has only positive eigenvalues.
By this notation we indicate the square -root modular operator  $\bm\Delta_{\tilde{\bf G},g}^{1/2}$ 
acting on the topological field $\ket{\Phi}$  as  
\bea\bm \Delta^{1/2}_{\tilde{\bf G},g}\ket{\Phi}&=&\phi\oplus {[G_{(1)}]}_{\mu\rho}g^{\rho\nu}\omega_{\nu}dx^{\mu}.
\eea
Using the flattened trace ($\mbox{Tr}_F$) formalism introduced in Ref. \cite{bianconi2025gravity} we obtain that the Lagrangian $\mathcal{L}$ indicating the GQRE can be expressed as \bea
\mathcal{L}=-\mbox{Tr}_F\ln \bm\Delta^{1/2}_{\tilde{\bf G},g}=-\mbox{Tr}\ln {\bf G}{g}^{-1},
\eea
where the trace $\mbox{Tr}$ of the last expression indicates the usual trace of a matrix.
Therefore the action $\mathcal{S}$ defined in Eq.(\ref{action}) and associated with the  GQRE Lagrangian given by Eq.(\ref{Lag00}), extends the definition of Araki~\cite{araki1975relative,ohya2004quantum} quantum relative entropy and provides geometrical definition of this entropy by treating metrics as quantum operators. 

{\bf The Perona-Malik algorithm as the gradient flow of the GfE action}
The GfE action, associated with the  Lagrangian given by Eq.(\ref{Lag}), expressing the GQRE between the metric   $g_{\mu\nu}=\eta_{\mu\nu}$ of the 2D support of the image and the induced metric ${ G}_{\mu\nu}$ defined in (\ref{induced}), is given by
\bea
\mathcal{S}=-\frac{1}{2}\int_\Omega d{\bf r} \ln(1+\alpha|\nabla \phi|^2).
\eea
Thus, the Perona-Malik algorithm can be easily shown to be the gradient flow that maximizes this action,  which is given by
\bea
\frac{d\phi({\bf r},t)}{dt}&=&\frac{\delta \mathcal{S}}{\delta \phi({\bf x},t)}.\eea
Indeed in this way we get the dynamical equations
\bea
\frac{d\phi({\bf r},t)}{dt}&=&\alpha\nabla_{\mu}\rho(|\nabla\phi|^2)\nabla^{\mu}\phi({\bf r},t),\eea
where $\rho(|
\nabla\phi|^2) $ is given by Eq.(\ref{rho}), i.e.
\bea
\rho(|\nabla\phi|^2)=\frac{1}{1+\alpha|\nabla\phi|^2}
\eea
with initial condition $\phi({\bf r},0)=\psi({\bf r}).$
Therefore, upon a rescaling of the time $t\to  t/\alpha$ we recover the Perona-Malik algorithm defined in Eq.(\ref{PM}).
As anticipated above, the GfE action, given by the GQRE between the metric $g$ and the metric induced by the image ${\bf G}$ provides the information theory principle that justifies the choice  of $\rho(|\nabla\phi|^2)$.

{\bf Conclusions.}
 {This work shows that the GfE approach that gives rise to modified gravity, provides  the entropic quantum gravity foundations of the  Perona-Malik algorithm for anisotropic diffusion.
 In particular, we have revealed that the Perona-Malik algorithm is the gradient flow that maximizes  the GfE action calculated among  the flat $2D$ metric $g_{\mu\nu}=\eta_{\mu\nu}$ and the metric $G_{\mu\nu}$ induced by the image.  This implies that the Perona-Malik algorithm can be interpreted as the outcome of the tension between  the metric of the flat $2D$ support of the image and the metric induced by the image on the $2D$ plane  trying to minimize their differences locally.
Interestingly, the maximization of the  GfE action is compatible with heterogeneous and complex images with sharp contours. This result is in sharp contrast with the  expectations arising from the classical maximum entropy principle and might be reflected also in the properties of the solutions of the GfE modified gravity equations. Therefore this result might indicate  relevant consequences of adopting the GQRE as the Lagrangian for both quantum gravity and  image processing algorithms.} 

 {From the point of view of artificial intelligence, our findings establish  solid quantum information-theoretic grounds for anisotropic diffusion and the Perona-Malik algorithm, justifying the  ad hoc choice for the functional expression of the metric adopted by Perona and Malik.
From the point of view of entropic quantum gravity, these findings provide a rather immediate application of the GfE approach to machine learning opening new perspectives on the development of the next generation of AI diffusion algorithms. Moreover, in this application, the $2D$ metric  associated with the image is flat and Euclidean and  remains unchanged during the learning of the image, while, going beyond the warm-up scenario, the GfE approach envisages that this metric is  associated to the curvature, and  can evolve in time.
 Therefore, future research could explore the full potential of GfE action to develop the next generation of AI algorithms.}
 
 {Based on this considerations, our expectations are that the GfE approach could provide  new perspectives on physics inspired artificial intelligence, unsupervised learning~\cite{schmarje2021survey} and brain research.
On one side,  these results   might lead  to the formulation of a new generation of geometric diffusion algorithms.
On the other side, they might inspire research at the interface between topological and geometrical learning and brain research~\cite{santos2019topological,zhou2018hyperbolic,citti2015gauge,citti2014neuromathematics}.
 which could potentially capture, among the other things, the main mechanisms beyond brain illusions such as the Kanizsa triangle \cite{kanizsa1976subjective}.
Thus the GfE approach might turn to be a fertile framework for proposing the next generation of  diffusion models fully based on the information encoded in the geometrical description of data.}

 {In summary,   this work demonstrates the common foundations of GfE and anisotropic diffusion for image processing and indicates that the full GfE action could potentially offer valuable insights for developing future generations of AI algorithms.}
\\
\section*{Acknowledgements}
The author acknowledges  interesting scientific discussions with Giovanna Citti and Alessandro Sarti and thanks them for pointing out  Ref. \cite{kimmel1997high}. This work was partially supported by a grant from the Simons Foundation. The author would like to thank the Isaac Newton Institute for Mathematical Sciences, Cambridge, for support and hospitality during the programme Hypergraphs: Theory and Applications, where work on this paper was undertaken. This work was supported by EPSRC grant EP/V521929/1.
\bibliographystyle{unsrt}
\bibliography{references}

\end{document}